\documentclass[letterpaper,twocolumn,aps,prl,floatfix,superscriptaddress,footinbib,preprintnumbers]{revtex4-2}
\usepackage{float}	
\usepackage[caption=false]{subfig}
\usepackage{graphicx}	
\usepackage{mathtools}
\usepackage{amsfonts}
\usepackage{bm}
\usepackage{slashed}
\usepackage{circuitikz}
\usepackage{import}
\usepackage[colorlinks=true]{hyperref}
\usepackage[capitalise]{cleveref}
\usepackage[title,titletoc]{appendix}
\usepackage{amssymb}

\newcommand{\sur}{{S \cup R}}
\newcommand{\Tr}{{\text{Tr}}}

\newcommand*{\ket}[1]{\lvert #1 \rangle}

\newcommand*{\ketbra}[2]{\lvert #1 \rangle\!\langle #2 \rvert}

\newcommand{\LParen}{ \bm{(} }
\newcommand{\RParen}{ \bm{)} }

\begin{document}

\preprint{UMD-PP-024-05, IQuS@UW-21-075}

\title{Quantum thermodynamics of nonequilibrium processes in lattice gauge theories}

\author{Zohreh~Davoudi}
\email{davoudi@umd.edu}
\affiliation{Department of Physics, University of Maryland, College Park, MD 20742 USA}
\affiliation{Maryland Center for Fundamental Physics, University of Maryland, College Park, MD 20742, USA}
\affiliation{Joint Center for Quantum Information and Computer Science, NIST and University of Maryland, College Park, MD 20742 USA}
\affiliation{The NSF Institute for Robust Quantum Simulation, University of Maryland, College Park, Maryland 20742, USA}

\author{Christopher~Jarzynski}
\email{cjarzyns@umd.edu }
\affiliation{Department of Chemistry and Biochemistry, University of Maryland, College Park, MD 20742 USA}
\affiliation{Institute for Physical Science and Technology, University of Maryland, College Park, MD 20742, USA}
\affiliation{Department of Physics, University of Maryland, College Park, MD 20742 USA}
\affiliation{The NSF Institute for Robust Quantum Simulation, University of Maryland, College Park, Maryland 20742, USA}

\author{Niklas~Mueller}
\email{niklasmu@uw.edu}
\affiliation{InQubator for Quantum Simulation (IQuS), Department of Physics,
University of Washington, Seattle, WA 98195, USA}

\author{Greeshma~Oruganti}
\email{gshivali@umd.edu }
\affiliation{Institute for Physical Science and Technology, University of Maryland, College Park, MD 20742, USA}
\affiliation{The NSF Institute for Robust Quantum Simulation, University of Maryland, College Park, Maryland 20742, USA}

\author{Connor~Powers}
\email{cdpowers@umd.edu}
\thanks{corresponding author.}
\affiliation{Department of Physics, University of Maryland, College Park, MD 20742 USA}
\affiliation{Maryland Center for Fundamental Physics, 
University of Maryland, College Park, MD 20742, USA}
\affiliation{Joint Center for Quantum Information and Computer Science, NIST and University of Maryland, College Park, MD 20742 USA}
\affiliation{The NSF Institute for Robust Quantum Simulation, University of Maryland, College Park, Maryland 20742, USA}

\author{Nicole~Yunger~Halpern}
\email{nicoleyh@umd.edu}
\affiliation{Joint Center for Quantum Information and Computer Science, NIST and University of Maryland, College Park, MD 20742 USA}
\affiliation{Institute for Physical Science and Technology, University of Maryland, College Park, MD 20742, USA}
\affiliation{The NSF Institute for Robust Quantum Simulation, University of Maryland, College Park, Maryland 20742, USA}

\begin{abstract}
A key objective in nuclear and high-energy physics is to describe nonequilibrium dynamics of matter, e.g., in the early universe and in particle colliders, starting from the Standard Model. Classical-computing methods, via the framework of lattice gauge theory, have experienced limited success in this mission. Quantum simulation of lattice gauge theories holds promise for overcoming computational limitations. Because of local constraints (Gauss’s laws), lattice gauge theories have an intricate Hilbert-space structure. This structure complicates the definition of thermodynamic properties of systems coupled to reservoirs during equilibrium and nonequilibrium processes. We show how to define thermodynamic quantities such as work and heat using strong-coupling thermodynamics, a framework that has recently burgeoned within the field of quantum thermodynamics. Our definitions suit instantaneous quenches, simple nonequilibrium processes undertaken in quantum simulators. To illustrate our framework, we compute the work and heat exchanged during a quench in a $\mathbb{Z}_2$ lattice gauge theory coupled to matter in 1+1 dimensions. The thermodynamic quantities, as functions of the quench parameter, evidence a phase transition. For general thermal states, we derive a simple relation between a quantum many-body system's entanglement Hamiltonian, measurable with quantum-information-processing tools, and the Hamiltonian of mean force, used to define strong-coupling thermodynamic quantities.
\end{abstract}

\maketitle

\emph{Introduction.}---An overarching goal in nuclear and high-energy physics is to simulate strongly interacting matter, starting from gauge theories.
Key focuses include nonequilibrium phenomena described by quantum chromomodynamics (QCD), the theory of the strong force. Nonequilibrium phenomena arise in ultrarelativistic particle collisions~\cite{vogt2007ultrarelativistic,florkowski2010phenomenology} and in the early universe~\cite{peebles1993principles,kolb2018early}. Theoretical studies~\cite{kogut2004the,ghiglieri2020perturbative,berges2021qcd,lovato2022long,achenbach2023present} of in- and out-of-equilibrium phases of QCD, and of its thermalization mechanisms, are often restricted to extreme parameter regimes, to facilitate perturbation theory. Alternatively, studies feature simple (often low-dimensional) models to capture qualitative features of QCD.

Studying QCD and other strongly interacting gauge theories requires nonperturbative tools, as enabled by 
lattice gauge theory (LGT)~\cite{wilson1974confinement,kogut1975hamiltonian,kogut1979introduction,creutz1983quarks,creutz1983monte}. Within the path-integral formulation of LGTs, Monte-Carlo simulations can be feasible if Euclidean (imaginary) time replaces Minkowski (real) time.
The scheme permits parallels with statistical mechanics: Euclidean time stands in for inverse temperature, and vacuum expectation values serve as thermal averages. LGT has enabled thermodynamic studies of the QCD equation of state at small chemical potentials~\cite{aoki2006order,aoki2006qcd,aoki2009qcd,borsanyi2010there,bhattacharya2014qcd,bazavov2012chiral,ding2015thermodynamics,ratti2018lattice,bazavov2019hot,philipsen2019constraining,guenther2021overview,nagata2022finite}. Nonetheless, the sampling weight in Monte-Carlo computations can become nonreal, requiring infeasibly many samples~\cite{troyer2005computational,nagata2022finite,gattringer2016approaches,alexandru2022complex,pan2022sign}. These limitations do not inhibit Hamiltonian-based approaches, such as tensor-network methods~\cite{banuls2018tensor,banuls2020review,meurice2022tensor} and quantum simulation~\cite{banuls2020simulating,klco2022standard,bauer2023quantumprx,bauer2023quantumnature, di2023quantum, halimeh2023cold}. These approaches, hence, suit thermodynamic studies of gauge theories, in and out of equilibrium. Still, we need a modern description of LGTs in the language of quantum thermodynamics.

The field of quantum thermodynamics extends conventional thermodynamics to small and quantum systems that exchange heat and work~\cite{goold2016role,vinjanampathy2016quantum,yungerhalpern2022quantum}. A typical setup features a subsystem of interest (the \emph{system}) interacting with a \emph{reservoir} of inaccessible degrees of freedom (DOFs)~\cite{breuer2007the,deffner2019quantum}. The coupling is often weak because a system and reservoir typically interact only at their shared boundary. This boundary is of lower dimensionality than the system, whose volume is proportional to its internal energy. Hence the interaction energy is much smaller than the system's and reservoir's internal energies. Yet quantum systems and reservoirs can be small while behaving thermodynamically~\cite{d2016quantum,kaufman2016quantum,polkovnikov2016thermalization}; their interactions need not be negligible. 

Hence the subfield of strong-coupling (quantum) thermodynamics has burgeoned recently~\cite{binder2015quantum,seifert2016first,strasberg2016nonequilibrium, jarzynski2017stochastic,miller2018hamiltonian,perarnau2018strong,strasberg2019non,rivas2020strong,strasberg2020thermodynamics,xu2023quantum}. Strong coupling blurs the system-reservoir boundary,
complicating the definition of the system's internal energy, work, and heat. Nonetheless, definitions have been developed and obey thermodynamic laws~\cite{Oruganti2024how,rivas2020strong,rivas2019refined,miller2018hamiltonian,campisi2009thermodynamics,strasberg2019non}.

These considerations raise the question \emph{can weak-coupling thermodynamics describe LGTs?} Local constraints, or Gauss's laws, form the defining feature of (lattice) gauge theories: the charge at a site balances the electric-field flux emanating from the site. Only a subspace of the gauge theory's Hilbert space satisfies Gauss's laws and consists of physical states. This restriction complicates the partitioning of the system into subsystems: the values of the fields on one side of any partition depend on the values on the other side. Yet a partitioning is used to define quantities such as the bipartite entanglement entropy~\cite{buividovich2008entanglement,donnelly2012decomposition,casini2014remarks,radicevic2014notes,aoki2015definition,soni2016aspects,van2016entanglement,bulgarelli2024duality}. The system-reservoir partitioning in LGTs, we posit, can resemble that in strong-coupling thermodynamics. Hence, we use strong-coupling thermodynamics to define thermodynamic properties of LGTs.

We answer the following questions: What are the work and heat exchanged during instantaneous quenches~\cite{mitra2018quantum} ---simple nonequilibrium processes created in quantum simulations~\cite{altman2021quantum,daley2022practical}? Can these quantities signal phase transitions~\footnote{Relationships between work distributions and phase transitions were explored recently in the context of spin systems~\cite{lin2023work,kiely2023entropy,zhang2023excited,zawadzki2020work,wang2017probing, mascarenhas2014work,mzaouali2021work,fei2020work}.}? Can such quantities be measured efficiently with quantum-information-processing tools? 

We describe how to compute internal-energy changes. We further show how to define work and heat consistently with the first and second laws of thermodynamics~\cite{Oruganti2024how}. A simple model illustrates this framework: a $\mathbb{Z}_2$ LGT coupled to hardcore bosonic matter in 1+1 dimensions (D). We observe that thermodynamic quantities, as functions of chemical potential,
signal an apparent phase transition. Furthermore, we bridge the fields of quantum information theory and strong-coupling quantum thermodynamics:
we show that the entanglement Hamiltonian~\cite{li2008entanglement,dalmonte2022entanglement}, which can often be efficiently measured  experimentally~\cite{kokail2021entanglement,kokail2021quantum,joshi2023exploring,mueller2023quantum}, is related to the Hamiltonian of mean force~\cite{miller2018hamiltonian}, which underpins strong-coupling quantum thermodynamics.
\begin{figure*}[t!]
    \centering
    \includegraphics[scale=0.315]{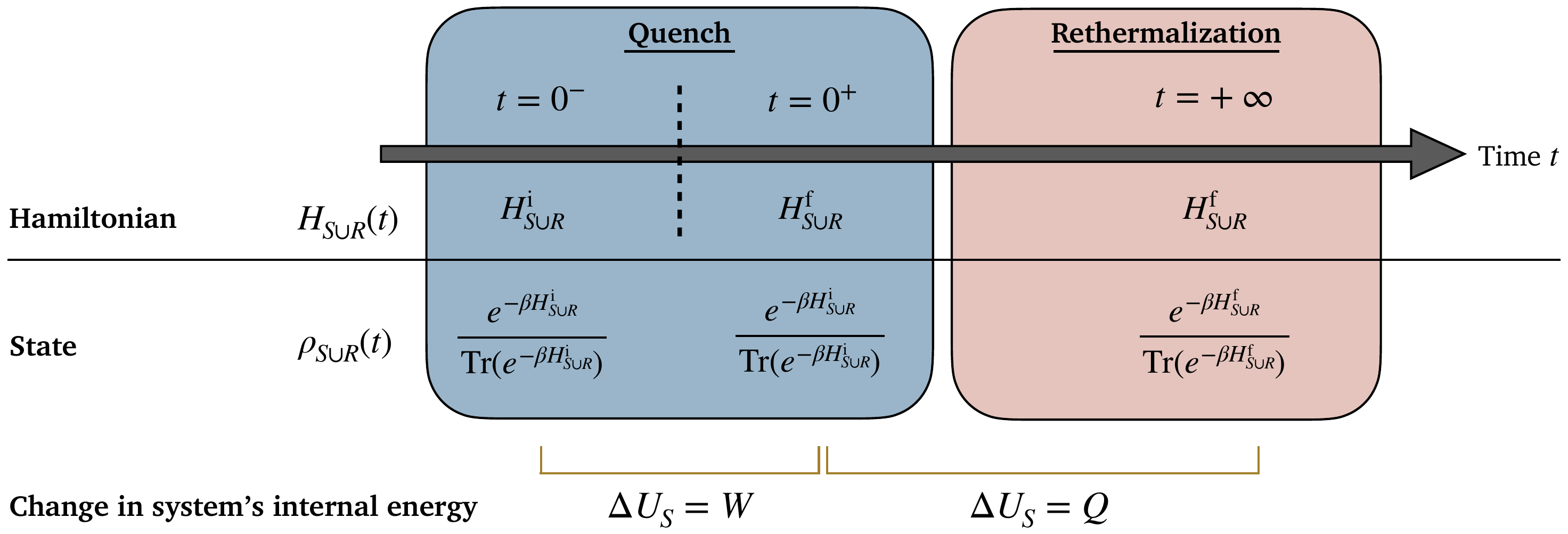}
    \caption{Overview of relevant quench protocol.
    The system starts in a global Gibbs state. 
    At $t=0$, the system Hamiltonian $H_S$ is quenched instantaneously. Under the new total Hamiltonian, the system-reservoir composite equilibrates to a global Gibbs state with the initial state's temperature.
    }
    \label{fig:quenchprotocol}
\end{figure*}
%

\emph{Review of strong-coupling quantum thermodynamics.}---Consider a system $S$ and a reservoir $R$. The composite $S \cup R$ evolves under the Hamiltonian 
\begin{equation}
\label{eq:genericHamiltonian}
    H_{S\cup R} \coloneqq H_S+H_R+V_{S\cup R}\, .
\end{equation}
$H_S$ and $H_R$ denote system's and reservoir's Hamiltonians, respectively; $V_{S\cup R}$ denotes the interaction. During a thermodynamic process, $S$ can absorb heat $Q$ and work $W$. \emph{Heat}, \emph{work}, and \emph{internal energy} refer to averages throughout the paper.

In weak-coupling quantum thermodynamics, $V_{S\cup R}$ contributes negligibly to the total internal energy, $U_{S\cup R} \coloneqq \langle H_{S\cup R} \rangle$. Hence the system's internal energy is $U_S \coloneqq \langle H_S\rangle$~\cite{rivas2019refined}. One can measure $U_S$ by accessing only system DOFs. In contrast, when $\langle V_{S\cup R}\rangle$ is comparable to $\langle H_S \rangle$, one must use strong-coupling thermodynamics.

How much $V_{S \cup R}$ contributes to $U_S$ is ambiguous. To resolve the ambiguity, one can define the global Gibbs state with respect to an inverse temperature $\beta$: $\pi_\sur \coloneqq e^{-\beta H_\sur} / Z_{\sur}$. The partition function is $Z_{\sur} \coloneqq \Tr (e^{-\beta H_\sur})$. Throughout this work, we denote thermal states by $\pi$ and general density matrices by $\rho$. $\Tr_X$ denotes the partial trace over $X$. Consider tracing out the reservoir from $\pi_\sur$.
The system's reduced density matrix is $\pi_S \coloneqq \Tr_R (\pi_\sur) \equiv e^{-\beta H_S^*}/Z_S^*$. This is a thermal state with respect to an effective Hamiltonian, \emph{Hamiltonian of mean force}~\cite{miller2018hamiltonian}: 
\begin{equation}\label{eq:HMF}
    H_S^* \coloneqq -\frac{1}{\beta} \ln \left( \frac{\Tr_R \left( e^{-\beta H_\sur} \right) }{Z_R}
    \right) \, ,
\end{equation}
wherein $Z_R \coloneqq \Tr_R(e^{-\beta H_R})$ and $Z_S^* \coloneqq \Tr_S(e^{-\beta H_S^*})=Z_\sur/Z_R$. Equation~\eqref{eq:HMF} underlies an intuitive definition for the system's free energy~\cite{miller2018hamiltonian,jarzynski2004nonequilibrium,talkner2020colloquium},
\begin{equation}
\label{eq:FS*-def}
    F_S \coloneqq -\frac{1}{\beta} \ln (Z_S^*)
    = U_S-\frac{\mathcal{S}}{\beta} \, .
\end{equation}
The second equality holds in equilibrium if $\mathcal{S} \coloneqq - \Tr (\pi_S \ln \pi_S)$ denotes the thermal state's von Neumann entropy~\footnote{Boltzmann's constant is set to one throughout the paper.}.

It is natural to equate $\langle H_S^* \rangle$ with $U_S$~\cite{jarzynski2004nonequilibrium,jarzynski2017stochastic,rivas2019refined,Oruganti2024how}. Consequently, one need not access reservoir DOFs to compute $U_S$. Any change in $U_S$ comes from work and heat, by the first law of thermodynamics: $\Delta U_S=W+Q$. By the second law, $\Delta F_S \leq W$, or $Q \leq \Delta \mathcal{S}/\beta$. Here, $\Delta U_S$, $\Delta F_S$ and $\Delta \mathcal{S}$ denote, respectively, net changes in the system's internal energy, free energy, and entropy. Intuition guides the separation of $\Delta U_S$ into $W$ and $Q$: work comes from changes in the system's Hamiltonian; and heat, from changes in the system's state~\cite{vinjanampathy2016quantum}. To define work and heat, we now specify a thermodynamic process.

\emph{Instantaneous quenches.}---During a \emph{quench}, the Hamiltonian changes rapidly. Quenches generate nonequilibrium conditions and are studied in diverse quantum-simulation experiments~\cite{eisert2015quantum,bernien2017probing,zhang2017observation,tan2021domain,ebadi2021quantum,de2023non,martinez2016real,klco2018quantum,nguyen2022digital,kharzeev2020real,zhou2022thermalization,mueller2023quantum}. Figure~\ref{fig:quenchprotocol} depicts the quench studied here. The system-reservoir composite begins in a Gibbs state: 
$\rho_{S\cup R}^\text{i} \coloneqq \rho_{S\cup R}(t{=}0^-)
= \pi_{S\cup R}^\text{i} \coloneqq e^{-\beta H_{S\cup R}^\text{i}} / {\rm{Tr} \big( e^{-\beta H_{S\cup R}^\text{i}} \big) }$, wherein the initial composite Hamiltonian $H_{S\cup R}^\text{i} \coloneqq H_{S\cup R}(t{=}0^-)$. Equation~\eqref{eq:genericHamiltonian}, with a time-dependent $H_S(t)$, specifies $H_\sur$. At time $t=0$, $H_S(t)$ is instantaneously quenched from $H_S(t{=}0^-) \coloneqq H_S^{\text{i}}$ to 
$H_S(t{=}0^+) \coloneqq H_S^{\text{f}}$. The system-reservoir composite equilibrates to a Gibbs state under the final total Hamiltonian $H_{S\cup R}(t{=}0^+) \coloneqq H_{S\cup R}^\text{f}$. The initial and final Gibbs states share the temperature $\beta^{-1}$ by assumption. Justifications include the possibility that $S\cup R$ couples weakly to a larger reservoir, at a temperature $\beta^{-1}$, in the distant past and future~\cite{Oruganti2024how}. 

Practical work and heat definitions depend on system DOFs alone. We present such definitions, for the quenches just described, in Ref.~\cite{Oruganti2024how}; see also Refs.~\cite{jarzynski2004nonequilibrium,seifert2016first,miller2018hamiltonian}. The system's internal energy is 
\begin{equation}
    U_S(t)\coloneqq \langle H_S^*(t) \rangle 
    = \Tr \bm{(} \rho_S(t) H_S^*(t) \bm{)} \, ,
\end{equation}
for arbitrary states $\rho_S(t) \coloneqq \text{Tr}_R\bm{(}\rho_{S \cup R}(t)\bm{)}$. $H_S^*$ becomes time-dependent under the replacement
$\text{Tr}_R \bm{(} e^{-\beta H_{S\cup R}} \bm{)} \mapsto \text{Tr}_R \bm{(} e^{-\beta H_{S\cup R}(t)} \bm{)}$ in Eq.~\eqref{eq:HMF}. During the quench, $\Delta U_S$ equals the work absorbed by $S$:
\begin{equation}
\label{eq:W-quench}
    W \coloneqq \Tr_S \bm{(} \rho^{\text{i}}_{S} H_S^*(t=0^+) \bm{)} -\Tr_S \bm{(} \rho^{\text{i}}_{S} H_S^*(t=0^-) \bm{)} \, .
\end{equation}
During the equilibration, $\Delta U_S$ equals the heat absorbed by $S$:
\begin{equation}
\label{eq:Q-quench}
    Q \coloneqq \Tr_S \LParen \rho^{\text{f}}_{S} H_S^*(t=0^+)\RParen 
    -\Tr_S \LParen \rho^{\text{i}}_{S} H_S^*(t=0^+) \RParen \, .
\end{equation}
These definitions are intuitive and obey the first and second laws of thermodynamics~\cite{Oruganti2024how}. Upon identifying a Hamiltonian of the form in Eq.~\eqref{eq:genericHamiltonian}, one calculates work and heat by measuring $\langle H_S^* \rangle$. We show next how to measure this quantity.

\emph{Measuring thermodynamic quantities in quantum simulations.}---Here, we derive a relation between strong-coupling-thermodynamics quantities and a quantity called the entanglement Hamiltonian. Every density matrix $\rho$ can be expressed as $\rho = \sum_k p_k \ketbra{k}{k}$. The $\ket{k}$ denote eigenstates; and the $p_k \in [0, 1]$, probabilities. Define $\lambda_k \coloneqq - \ln (p_k) \geq 0$, such that 
$\rho = \sum_k e^{-\lambda_k} \ketbra{k}{k}$. This expansion has the form of a thermal state at unit temperature (with the normalization factor, or partition function, absorbed into the $e^{-\lambda_k}$). For $\rho=\rho_S \coloneqq \rm{Tr}_R(\rho_{S\cup R})$, this Hamiltonian is the \emph{(bipartite) entanglement Hamiltonian}~\cite{li2008entanglement,dalmonte2022entanglement}, 
\begin{equation}
H^{\rm{ent}}_{S} \coloneqq -\ln (\rho_S)\,.
\label{eq:def-H-ent}
\end{equation}
This operator contains more information than the bipartite entanglement entropy. It has spawned numerous studies in quantum information theory and many-body physics.~\footnote{For instance, the $\lambda_k$ statistics imply whether a system behaves ergodically~\cite{geraedts2016many,yang2017entanglement,chang2019evolution,rakovszky2019signatures,mueller2022thermalization}. (Ergodicity is necessary for thermalization.) Also, a gap in the $H_S^{\rm ent}$ spectrum may signal a topologically ordered phase~\cite{li2008entanglement,mueller2022thermalization,zache2022entanglement}.} Parameterized \emph{Ans\"atze} for entanglement Hamiltonians~\cite{dalmonte2022entanglement}, with random-measurement protocols~\cite{elben2019statistical,huang2020predicting,huang2022quantum,elben2023randomized}, enable tomography of ground and nonequilibrium states~\cite{pichler2016measurement,dalmonte2018quantum,kokail2021entanglement,kokail2021quantum,joshi2023exploring}, including of LGTs~\cite{mueller2023quantum,bringewatt2023randomized,mueller2024notes}.
\begin{figure*}[t]
    \centering
    \includegraphics[scale=0.285]{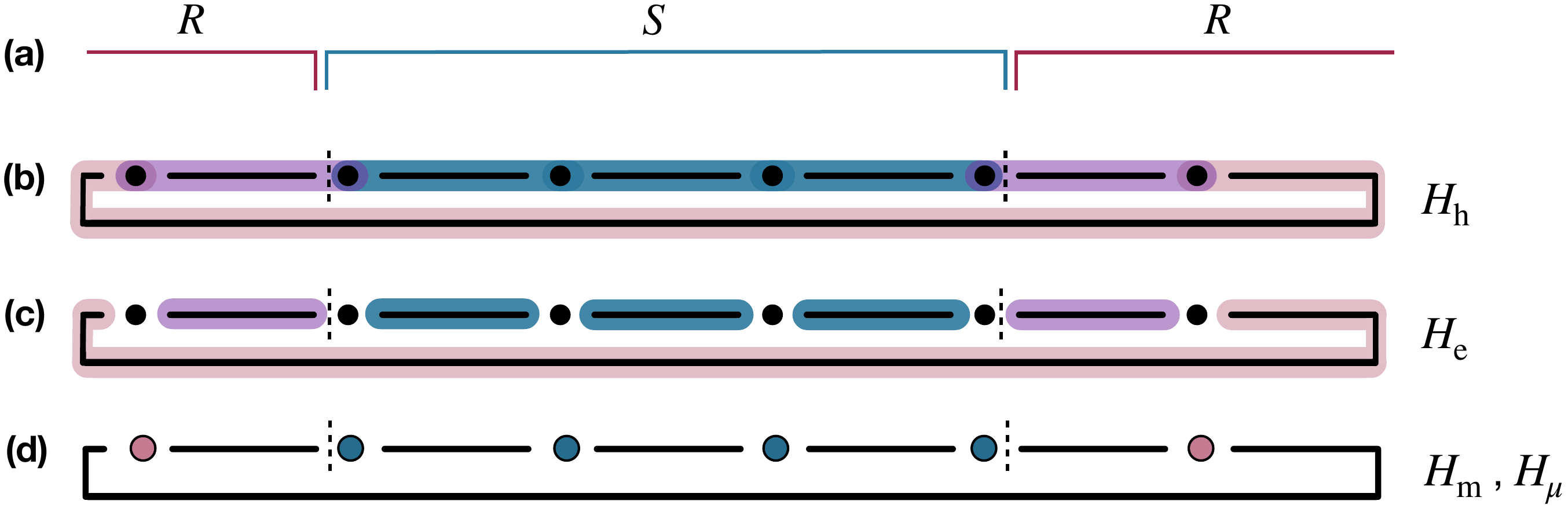}
    \caption{ 
    (a) Partitioning of DOFs into system ($S$) and reservoir ($R$) DOFs. Partitioning of total Hamiltonian's (b) hopping terms, (c) electric-field terms, and (d) mass and chemical-potential terms into $H_S(t)$ (teal), $H_R$ (light red), and $V_{S\cup R}$ (purple). Matter (gauge-field) DOFs reside on lattice sites (links).  The last link wraps around to indicate the lattice's periodicity.
    }
    \label{fig:1D lattice partitioning}
\end{figure*}

To leverage  such tomography tools, we prove a relation between $H^{\rm{ent}}_{S}$ and $H_S^*$ for thermal states. We rewrite the Hamiltonian of mean force as 
\begin{subequations}
\begin{align}
\label{eq:H*-I}
    H_S^* &= -\frac{1}{\beta} \ln \left( 
    \frac{\Tr_R(e^{-\beta H_\sur})}{
          \Tr_R(e^{-\beta H_R})}\right) \\
    \label{eq:H*-III}
    &= -\frac{1}{\beta} \ln \left (\rho_S \frac{Z_\sur}{Z_R}\right ) \,.
\end{align}
\end{subequations}
The system's reduced state is $\rho_S=\pi_S \coloneqq \Tr_R(e^{-\beta H_\sur})/Z_\sur$. Using Eqs.~(\ref{eq:def-H-ent}) and (\ref{eq:FS*-def}) yields the relation between the entanglement Hamiltonian and the Hamiltonian of mean force:
\begin{equation}
\label{eq:EH-HMF_relation}
    H_S^* = \frac{1}{\beta} H^{\rm{ent}}_{S} + F_S \, .
\end{equation}
$\mathbb{I}_S$ denotes the identity operator defined on $S$.

One can measure the first term in Eq.~\eqref{eq:EH-HMF_relation} using the aforementioned tomography tools (if the \emph{Ans\"atze} model $H^{\rm{ent}}_{S}$ accurately).  
To measure $F_S$, one must measure $Z_{S \cup R}$ and $Z_R$~\cite{matsumoto2022calculation,wu2022estimating,arunachalam2022simpler,bravyi2021complexity}, necessitating access to reservoir DOFs. To avoid measuring the second term in \cref{eq:EH-HMF_relation}, we study the average dissipated work, $ W_{\rm{diss}} \coloneqq W - \Delta F_S$, a measure of entropy production~\cite{kawai2007dissipation,jarzynski2011equalities}. For our quench protocol,
\begin{equation}
\label{eq:dissipated_work}
    W_{\rm{diss}} =\frac{1}{\beta} 
    \Tr_S \bm{(} \rho^{\text{i}}_{S} 
    \left[ H^{\rm{ent}}_{S}(t{=}0^+) - H^{\rm{ent}}_{S}(t{=}0^-) \right] \bm{)} \, .
\end{equation}
Hence, one can infer $W_{\rm{diss}}$ upon measuring the entanglement Hamiltonian alone.

\emph{Lattice gauge theories and strong-coupling quantum thermodynamics.}---We use the Hamiltonian formulation of LGTs~\cite{kogut1979introduction}, which suits quantum simulation. As matter and gauge fields undergo local symmetry-group transformations, the observables remain invariant. The symmetry restricts the states to a \emph{physical subspace}: for each site $n$, a \emph{Gauss-law} operator $G_n$ acts on $n$ and commutes with the Hamiltonian. The full Hilbert space (spanned by eigenbases of electric and matter fields) decomposes into $G_n$ eigenspaces, for each $n$. The eigenspace labeled by some eigenvalue $g$, and shared by all the $G_n$, is the physical subspace!\footnote{In the Standard Model of particle physics, nature sets the value of $g$.}. That is, for any physical state $|\Psi_{\rm{phys}}\rangle$, 
\begin{equation}
\label{eq:physicality}
    G_n |\Psi_{\rm{phys}}\rangle = g|\Psi_{\rm{phys}}\rangle \, , \quad \forall n \, .
\end{equation}
In electrodynamics, $G_n = \bm{\nabla}\cdot \bm{E}_n - \rho_n$. $\bm{E}_n$ denotes the electric field, and $\rho_n$ denotes the (dynamical) electric-charge density, both at site $n$. Gauss's law follows from setting $g=0$ ($g \neq 0$) in \cref{eq:physicality} in the absence (presence) of a background static electric charge.

One may impose Gauss's laws by manually removing the unphysical states from the full Hilbert space~\cite{stryker2019oracles,raychowdhury2020solving}. Alternatively, the Hamiltonian $H_{S\cup R}$ may be replaced with $H_{S\cup R} + \sum_n f(G_n)$. $f(G_n)$ denotes a function of Gauss-law operators. Chosen properly, it penalizes transitions to unphysical states~\cite{zohar2011confinement,banerjee2012atomic,zohar2013simulating,tagliacozzo2013simulation,hauke2013quantum,davoudi2023towards,halimeh2021gauge,halimeh2022stabilizing}. Consider partitioning a lattice into a system $S$ and a reservoir $R$. Some Gauss-law penalty terms act on both $S$ and $R$: Gauss-law operators are multibody operators consisting of gauge and matter fields. Such penalty terms, thus, contribute to the $V_\sur$ in Eq.~\eqref{eq:genericHamiltonian}. Their contribution must be large to constrain the state to the physical subspace. Therefore, one cannot generally neglect the internal energy's dependence on $V_\sur$ when computing thermodynamic quantities; see Supplemental Material (SM)~\cite{SupplementalMaterial}. Consequently, LGTs can be described within the framework of strong-coupling thermodynamics.

\emph{Example of $\mathbb{Z}_2$ LGT coupled to matter in (1+1)D.}---Consider a $\mathbb{Z}_2$ gauge field (hardcore bosons) coupled to matter fields (chosen to be hardcore bosons). The initial state evolves under the Hamiltonian
\begin{equation}
\label{eq:1DHamiltonian}
\begin{split}
    H&= H_\mathrm{h} + H_\mathrm{e} + H_\mathrm{m} + H_\mathrm{\mu} + H_c \\ 
    &\equiv -J\sum_{n=0}^{N-1} \left( \sigma^+_n \tilde{\sigma}^z_n \sigma^-_{n+1} + {\rm{h.c.}} \right) 
    - \epsilon \sum_{n=0}^{N-1} \tilde{\sigma}^x_n \\& \quad + m\sum_{n=0}^{N-1} (-1)^n \sigma^+_n \sigma^-_n -  \sum_{n=0}^{N-1} \mu_n \sigma^+_n \sigma^-_n + c\sum_{n=0}^{N-1}\mathbb{I}_n\, ,
\end{split}
\end{equation}
on a one-dimensional $N$-site spatial lattice with periodic boundary conditions ($\sigma_{N}=\sigma_0$). $H_{\rm h}$, $H_{\rm e}$, $H_{\rm m}$, and $H_\mu$ represent the matter-hopping, electric-field, matter-mass, and matter-chemical-potential terms, respectively.~\footnote{The mass term's sign alternates between even-index and odd-index sites, consistently with the staggering of fermionic matter~\cite{kogut1975hamiltonian}. Despite working with a bosonic theory, we keep the staggering convention.} A constant $H_c$ is added such that $H_S(t)$, $H_R$, and $H_\sur(t)$ have only non-negative eigenvalues. $J$, $\epsilon$, $m$, and $\mu_n$ denote the hopping strength, electric-field strength, matter mass, and site-$n$ chemical potential, respectively. Pauli operator $\sigma_n$ acts on the Hilbert space of the site-$n$ matter field. Pauli operator $\tilde{\sigma}_n$ acts on the Hilbert space of the gauge field rightward of $n$. Specifically, $\tilde{\sigma}_n^x$ ($\tilde{\sigma}_n^z$) denotes the electric-field (gauge-link) operator.

The $\mathbb{Z}_2$ gauge transformation is generated by the Gauss-law operator~\footnote{
For a pure gauge theory, which lacks matter fields, the operator reduces to $G_n\coloneqq \tilde{\sigma}^x_n \tilde{\sigma}^x_{n-1}$. This is the familiar generator of a local $\mathbb{Z}_2$ symmetry.} 
%
\begin{equation}
\label{eq:Gausslawop}
    G_n = \tilde{\sigma}^x_n \tilde{\sigma}^x_{n-1}
    \exp \left( i \pi 
    \left[ \sigma^+_n \sigma^-_n + \frac{(-1)^n -1}{2} \right] \right) \, .
\end{equation}
The physical states obey Eq.~\eqref{eq:physicality} with $g=1$~\footnote{$G_n$ has only two eigenvalues, $\pm 1$. 
Physically, when $g=1$, if no matter (or antimatter) occupies site $n$, the electric field on the left equals that on the right. If site $n$ contains matter (or antimatter), the electric-field eigenvalues on the left and right have opposite signs.}. One can realize the gauge-invariant dynamics by adding $\sum_n f(G_n)=\kappa \sum_n(1-G_n)$ to $H$, then taking the limit $\kappa \rightarrow \infty$~\footnote{Numerical computations of this work are performed by manually restricting to the physical Hilbert space.}.

The lattice can be partitioned into a system $S$ and a reservoir $R$, as in Fig.~\ref{fig:1D lattice partitioning}(a). Also, the Hamiltonian~\eqref{eq:1DHamiltonian} decomposes as in Eq.~\eqref{eq:genericHamiltonian}. Some hopping terms act only on sites in $S$ ($R$) and so belong in $H_S$ ($H_R$). Other terms describe hopping between a site in $S$ and a site in $R$. These interaction terms belong in $V_{S\cup R}$; see Fig.~\ref{fig:1D lattice partitioning}(b). The electric-field Hamiltonian can be partitioned as follows; see Fig.~\ref{fig:1D lattice partitioning}(c). Terms acting on links in $S$ belong in $H_S(t)$. Terms acting on links in $R$, but touching the $S$-$R$ boundary, belong in $V_{S\cup R}$ due to Gauss's laws. Terms acting on links elsewhere in $R$ belong in $H_R$. Terms in $H_\text{m}$ and $H_\mu$ belong in $H_S(t)$ and $H_R$, depending on whether they act on a site in $S$ or $R$; see Fig. \ref{fig:1D lattice partitioning}(d). Finally, if penalty terms enforcing Gauss's law act at the boundary, they belong in $V_\sur$. Otherwise, they belong in $H_S$ or $H_R$. 

\begin{figure}[t]
    \centering
    \includegraphics[width=0.99\linewidth]{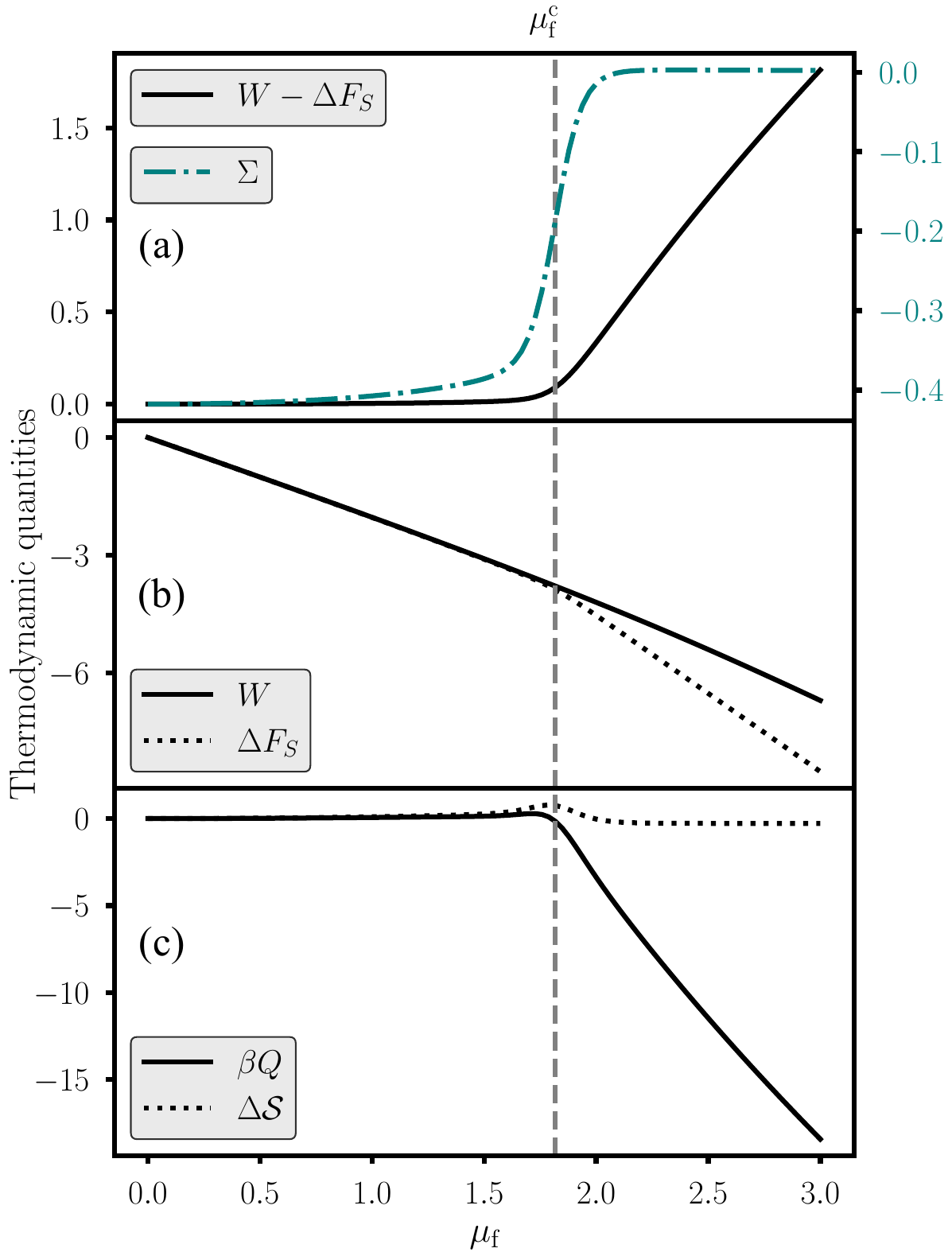}
    \caption{Thermodynamic quantities as functions of the final chemical potential, $\mu_{\text{f}}$, during the instantaneous quench 
    from $\mu_{\text{i}}=0$ with $N=6$, $N_S=4$, $t=-1/2$, $\epsilon=1/2$, $m=1/2$, $\mu_\text{i}=0$, and $\beta=10$. (a) Dissipated work $W_{\rm{diss}}$ (solid, black) and chiral condensate $\Sigma$ (dot-dashed, teal). (b) Work $W$ (solid, black) and change $\Delta F_S$ in the system's free energy (dotted, black). (c) Normalized heat $\beta Q$ (solid, black) and change $\Delta \mathcal{S}$ in the system's entropy (dotted, black). We also plot $\mu_\text{f}^{\rm c}$, the $\mu_{\text{f}}$ value where $\Sigma'$ is maximized (dashed, gray).
    }
    \label{fig:1D_muquench_fig}
\end{figure}

We set reservoir's chemical potential to zero and quench the chemical potential from $\mu_n = \mu_{\rm{i}}=0$ to $\mu_n=\mu_{\rm{f}} > 0$ at all sites $n$ in the system. We further define the \emph{chiral condensate}~\footnote{
$\Sigma$ has the form that the chiral condensate would have if the gauge field were coupled to staggered fermions~\cite{kogut1975hamiltonian} (a Jordan-Wigner transformation maps the fermion operators to Pauli operators). 
Such a chiral condensate serves as an order parameter for chiral phase transitions.
}
\begin{equation}
   \label{eq_chiral_cond}
   \Sigma \coloneqq \frac{1}{N_S} \sum_{n=0}^{N_S-1} (-1)^n 
   \langle \sigma^+_n \sigma^-_n 
   \rangle \, .
\end{equation}
$N_S$ denotes the system size. The expectation value is in the final thermal state, whose $\mu_n=\mu_\text{f}$ at all sites $n$ in $S$. If $\mu_\text{f} \gg \epsilon,m$, the system's 
state is dominated by the matter fields' all-spin-up state, yielding $\Sigma=0$. At other $\mu_\text{f}$ values, $\Sigma$ could be nonvanishing. In fact, $\Sigma$ suddenly changes from finite to vanishing values as $\mu_\text{f}$ increases, evidencing a phase transition, as Fig. \ref{fig:1D_muquench_fig}(a) shows. This apparent phase transition does not look perfectly sharp, due to finite-temperature (see SM~\cite{SupplementalMaterial}) and finite-size effects~\footnote{We have not performed a large-system analysis to confirm if the apparent phase transition persists in the thermodynamic limit. However, large-system analyses of a similar model (with the same discrete chiral symmetry as ours) reveal two phases, confined and deconfined~\cite{kebrivc2021confinement,kebrivc2023confinement}. While these studies do not directly probe the spontaneous breaking of the discrete chiral symmetry, the symmetry is broken in the confined phase.}. The critical value $\mu_\text{f} = \mu_\text{f}^{\rm c}$ denotes the transition point, where $\Sigma' \coloneqq d\Sigma/d\mu_\text{f}$ 
is maximized. 

Figure~\ref{fig:1D_muquench_fig} displays thermodynamic quantities, calculated as functions of $\mu_\text{f}$: $W_{\rm diss}$, $W$, $Q$,  $\Delta F_S$, and $\Delta \mathcal{S}$. As $\mu_\text{f}$ grows, $W_{\rm{diss}} \coloneqq W-\Delta F_S$ remains near zero until around the $\mu_\mathrm{f}$ value where the apparent phase transition occurs. Afterward, $W_{\rm diss}$ increases. Hence $W_{\rm diss}$ indicates the suspected phase transition clearly. Similarly, $W$ ($\beta Q$) begins deviating from $\Delta F_S$ ($\Delta \mathcal{S}$) around the transition.

Strong-coupling relations are valid in this model: $|\langle V_{S \cup R} \rangle / \langle H_S \rangle|$ is non-negligible ($\gtrsim 0.1$) in the initial and final states at all the $\mu_f$ values used. Nonetheless, in the SM~\cite{SupplementalMaterial}, we apply weak-coupling relations for comparison: $U_S=\langle H_S \rangle$ and $\rho_S =  e^{-\beta H_S}/\text{Tr}(e^{-\beta H_S})$. We find that the thermodynamic quantities' values differ from their strong-coupling counterparts: the weak-coupling quantities change less sharply, signaling the transition less reliably.

\emph{Outlook.}---This work shows how strong-coupling quantum thermodynamics applies to lattice gauge theories in and out of equilibrium.
The work further shows how entanglement Hamiltonians can be leveraged to measure thermodynamic quantities in quantum simulations. Our framework may be applied to explore further questions in the quantum thermodynamics of gauge theories. Examples include whether thermodynamic quantities signal topological~\cite{wen1990topological,kitaev2003fault,levin2005string,li2008entanglement,bernevig2013topological,seo2013thermodynamic,kempkes2016universalities,caputo2018topological,carollo2020geometry} or dynamical~\cite{heyl2018dynamical} phase transitions. While quench protocols in LGTs have been implemented in experiments~\cite{banerjee2012atomic,martinez2016real, klco2018quantum, schweizer2019floquet, yang2020observation, notarnicola2020real, surace2020lattice, klco20202, ciavarella2021trailhead, nguyen2022digital, de2022quantum,  mildenberger2022probing, halimeh2022tuning, zhou2022thermalization,mueller2023quantum,zhang2023observation,charles2024simulating}, a longer-term vision is to simulate more-general processes, including quantum-adiabatic ones, and particle collisions relevant to nuclear and high-energy physics~\cite{surace2021scattering,belyansky2023high,chai2023entanglement,su2024cold,farrell2024quantum,davoudi2024scattering}. Also, non-Abelian and higher-dimensional gauge theories merit studying within our framework. Developing quantum-simulation protocols for studying gauge-theory thermodynamics is an active frontier, and this work furthers this goal.

\emph{Acknowledgments.}---Z.~D., G.~O., and C.~P. were supported by the National Science Foundation (NSF) Quantum Leap Challenge Institutes (QLCI) (award no. OMA-2120757).
Z.D. further acknowledges funding by the Department of Energy (DOE), Office of Science, Early Career Award (award no. DESC0020271), and by the Department of Physics, Maryland Center for Fundamental Physics, and the College of Computer, Mathematical, and Natural Sciences at the University of Maryland, College Park. She is grateful for the hospitality of Perimeter Institute where part of this work was carried out. Research at Perimeter Institute is supported in part by the Government of Canada through the Department of Innovation, Science, and Economic Development, and by the Province of Ontario through the Ministry of Colleges and Universities. Z.D. was also supported in part by the Simons Foundation through the Simons Foundation Emmy Noether Fellows Program at Perimeter Institute.
C.~J. and N.~Y.~H. further acknowledge support from John Templeton Foundation (award no. 62422). N.~Y.~H. thanks Harry Miller for conversations about strong-coupling thermodynamics.
N.~M. acknowledges funding by the DOE, Office of Science, Office of Nuclear Physics, \href{https://iqus.uw.edu}{Inqubator for Quantum Simulation} (award no. DE-SC0020970).
C. P. is grateful for discussions on thermal gauge theories with Robert Pisarski, and for the hospitality of Brookhaven National Laboratory which hosted C.~P. as part of an Office of Science Graduate Student Research Fellowship.
G.~O. further acknowledges support from the American Association of University Women through an International Fellowship. 
Finally, N.~M., G.~O., C.~P., and N.~Y.~H. thank the participants of the InQubator for Quantum Simulation (IQuS) workshop ``\href{https://iqus.uw.edu/events/iqus-workshop-thermalization/}{Thermalization, from Cold Atoms to Hot Quantum Chromodynamics}'' at the University of Washington in September 2023 for many valuable discussions.

\bibliographystyle{h-physrev}
\bibliography{bibi.bib}

\begin{thebibliography}{100}

\bibitem{vogt2007ultrarelativistic}
R.~Vogt,
\newblock {\em Ultrarelativistic Heavy-Ion Collisions} (Elsevier, New York,
  2007).

\bibitem{florkowski2010phenomenology}
W.~Florkowski,
\newblock {\em Phenomenology of Ultra-Relativistic Heavy-Ion Collisions} (World
  Scientific Publishing Company, Singapore, 2010).

\bibitem{peebles1993principles}
P.~J.~E. Peebles,
\newblock {\em Principles of Physical Cosmology} (Princeton University Press,
  Princeton, New Jersey, 1993).

\bibitem{kolb2018early}
E.~Kolb,
\newblock {\em The Early Universe} (CRC Press, Boca Raton, Florida, 2018).

\bibitem{kogut2004the}
J.~B. Kogut and M.~A. Stephanov,
\newblock {\em {The Phases of Quantum Chromodynamics: From Confinement to
  Extreme Environments}} (Cambridge University Press, {Cambridge, England},
  2003).

\bibitem{ghiglieri2020perturbative}
J.~Ghiglieri, A.~Kurkela, M.~Strickland, and A.~Vuorinen,
\newblock Physics Reports {\bf 880}, 1 (2020).

\bibitem{berges2021qcd}
J.~Berges, M.~P. Heller, A.~Mazeliauskas, and R.~Venugopalan,
\newblock Reviews of Modern Physics {\bf 93}, 035003 (2021).

\bibitem{lovato2022long}
A.~Lovato {\em et~al.},
\newblock arXiv preprint arXiv:2211.02224  (2022).

\bibitem{achenbach2023present}
P.~Achenbach {\em et~al.},
\newblock Nuclear Physics A {\bf 1047}, 122874 (2024).

\bibitem{wilson1974confinement}
K.~G. Wilson,
\newblock Physical review D {\bf 10}, 2445 (1974).

\bibitem{kogut1975hamiltonian}
J.~Kogut and L.~Susskind,
\newblock Physical Review D {\bf 11}, 395 (1975).

\bibitem{kogut1979introduction}
J.~B. Kogut,
\newblock Reviews of Modern Physics {\bf 51}, 659 (1979).

\bibitem{creutz1983quarks}
M.~Creutz,
\newblock {\em Quarks, Gluons and Lattices} (Cambridge University Press,
  Cambridge, England, 1983).

\bibitem{creutz1983monte}
M.~Creutz, L.~Jacobs, and C.~Rebbi,
\newblock Physics Reports {\bf 95}, 201 (1983).

\bibitem{aoki2006order}
Y.~Aoki, G.~Endr{\H{o}}di, Z.~Fodor, S.~D. Katz, and K.~K. Szab{\'o},
\newblock Nature {\bf 443}, 675 (2006).

\bibitem{aoki2006qcd}
Y.~Aoki, Z.~Fodor, S.~Katz, and K.~K. Szab{\'o},
\newblock Physics Letters B {\bf 643}, 46 (2006).

\bibitem{aoki2009qcd}
Y.~Aoki {\em et~al.},
\newblock Journal of High Energy Physics {\bf 2009}, 088 (2009).

\bibitem{borsanyi2010there}
S.~Bors\'{a}nyi {\em et~al.},
\newblock Journal of High Energy Physics {\bf 2010}, 73 (2010).

\bibitem{bhattacharya2014qcd}
HotQCD Collaboration, T.~Bhattacharya {\em et~al.},
\newblock Phys. Rev. Lett. {\bf 113}, 082001 (2014).

\bibitem{bazavov2012chiral}
A.~Bazavov {\em et~al.},
\newblock Physical Review D {\bf 85}, 054503 (2012).

\bibitem{ding2015thermodynamics}
H.-T. Ding, F.~Karsch, and S.~Mukherjee,
\newblock International Journal of Modern Physics E {\bf 24}, 1530007 (2015).

\bibitem{ratti2018lattice}
C.~Ratti,
\newblock Reports on Progress in Physics {\bf 81}, 084301 (2018).

\bibitem{bazavov2019hot}
{USQCD Collaboration}, A.~Bazavov, F.~Karsch, S.~Mukherjee, and P.~Petreczky,
\newblock The European Physical Journal A {\bf 55}, 194 (2019).

\bibitem{philipsen2019constraining}
O.~Philipsen,
\newblock arXiv preprint arXiv:1912.04827  (2019).

\bibitem{guenther2021overview}
J.~N. Guenther,
\newblock The European Physical Journal A {\bf 57}, 136 (2021).

\bibitem{nagata2022finite}
K.~Nagata,
\newblock Progress in Particle and Nuclear Physics {\bf 127}, 103991 (2022).

\bibitem{troyer2005computational}
M.~Troyer and U.-J. Wiese,
\newblock Physical Review Letters {\bf 94}, 170201 (2005).

\bibitem{gattringer2016approaches}
C.~Gattringer and K.~Langfeld,
\newblock International Journal of Modern Physics A {\bf 31}, 1643007 (2016).

\bibitem{alexandru2022complex}
A.~Alexandru, G.~Ba{\c{s}}ar, P.~F. Bedaque, and N.~C. Warrington,
\newblock Reviews of Modern Physics {\bf 94}, 015006 (2022).

\bibitem{pan2022sign}
G.~Pan and Z.~Y. Meng,
\newblock arXiv preprint arXiv:2204.08777  (2022).

\bibitem{banuls2018tensor}
M.~C. Ba\~{n}uls, K.~Cichy, J.~I. Cirac, K.~Jansen, and S.~K{\"u}hn,
\newblock arXiv preprint arXiv:1810.12838  (2018).

\bibitem{banuls2020review}
M.~C. Ba\~{n}uls and K.~Cichy,
\newblock Reports on Progress in Physics {\bf 83}, 024401 (2020).

\bibitem{meurice2022tensor}
Y.~Meurice, R.~Sakai, and J.~Unmuth-Yockey,
\newblock Reviews of Modern Physics {\bf 94}, 025005 (2022).

\bibitem{banuls2020simulating}
M.~C. Ba\~{n}uls {\em et~al.},
\newblock The European Physical Journal D {\bf 74}, 165 (2020).

\bibitem{klco2022standard}
N.~Klco, A.~Roggero, and M.~J. Savage,
\newblock Reports on Progress in Physics {\bf 85}, 064301 (2022).

\bibitem{bauer2023quantumprx}
C.~W. Bauer {\em et~al.},
\newblock PRX Quantum {\bf 4}, 027001 (2023).

\bibitem{bauer2023quantumnature}
C.~W. Bauer, Z.~Davoudi, N.~Klco, and M.~J. Savage,
\newblock Nature Rev. Phys. {\bf 5}, 420 (2023).

\bibitem{di2023quantum}
A.~Di~Meglio {\em et~al.},
\newblock PRX Quantum {\bf 5}, 037001 (2024).

\bibitem{halimeh2023cold}
J.~C. Halimeh, M.~Aidelsburger, F.~Grusdt, P.~Hauke, and B.~Yang,
\newblock arXiv preprint arXiv:2310.12201  (2023).

\bibitem{goold2016role}
J.~Goold, M.~Huber, A.~Riera, L.~del Rio, and P.~Skrzypczyk,
\newblock Journal of Physics A: Mathematical and Theoretical {\bf 49}, 143001
  (2016).

\bibitem{vinjanampathy2016quantum}
S.~Vinjanampathy and J.~Anders,
\newblock Contemporary Physics {\bf 57}, 545 (2016).

\bibitem{yungerhalpern2022quantum}
N.~Yunger~Halpern,
\newblock {\em Quantum Steampunk: The Physics of Yesterday's Tomorrow} (JHU
  Press, Baltimore, 2022).

\bibitem{breuer2007the}
H.-P. Breuer and F.~Petruccione,
\newblock {\em The Theory of Open Quantum Systems} (Oxford U. Press, Oxford,
  2007).

\bibitem{deffner2019quantum}
S.~Deffner and S.~Campbell,
\newblock The principles of modern thermodynamics,
\newblock in {\em Quantum Thermodynamics}, 2053-2571, pp. 1--1 to 2--37, Morgan
  \& Claypool Publishers, San Rafael, 2019.

\bibitem{d2016quantum}
L.~D'Alessio, Y.~Kafri, A.~Polkovnikov, and M.~Rigol,
\newblock Advances in Physics {\bf 65}, 239 (2016).

\bibitem{kaufman2016quantum}
A.~M. Kaufman {\em et~al.},
\newblock Science {\bf 353}, 794 (2016).

\bibitem{polkovnikov2016thermalization}
A.~Polkovnikov and D.~Sels,
\newblock Science {\bf 353}, 752 (2016).

\bibitem{binder2015quantum}
F.~Binder, S.~Vinjanampathy, K.~Modi, and J.~Goold,
\newblock Physical Review E {\bf 91}, 032119 (2015).

\bibitem{seifert2016first}
U.~Seifert,
\newblock Phys. Rev. Lett. {\bf 116}, 020601 (2016).

\bibitem{strasberg2016nonequilibrium}
P.~Strasberg, G.~Schaller, N.~Lambert, and T.~Brandes,
\newblock New Journal of Physics {\bf 18}, 073007 (2016).

\bibitem{jarzynski2017stochastic}
C.~Jarzynski,
\newblock Phys. Rev. X {\bf 7}, 011008 (2017).

\bibitem{miller2018hamiltonian}
H.~J.~D. Miller,
\newblock {\em Hamiltonian of Mean Force for Strongly-Coupled Systems}
  (Springer International Publishing, Cham, 2018), pp. 531--549.

\bibitem{perarnau2018strong}
M.~Perarnau-Llobet, H.~Wilming, A.~Riera, R.~Gallego, and J.~Eisert,
\newblock Physical Review Letters {\bf 120}, 120602 (2018).

\bibitem{strasberg2019non}
P.~Strasberg and M.~Esposito,
\newblock Phys. Rev. E {\bf 99}, 012120 (2019).

\bibitem{rivas2020strong}
{\'A}.~Rivas,
\newblock Phys. Rev. Lett. {\bf 124}, 160601 (2020).

\bibitem{strasberg2020thermodynamics}
P.~Strasberg,
\newblock Quantum {\bf 4}, 240 (2020).

\bibitem{xu2023quantum}
Y.-Y. Xu, J.~Gong, and W.-M. Liu,
\newblock arXiv preprint arXiv:2304.08268  (2023).

\bibitem{Oruganti2024how}
Z.~Davoudi {\em et~al.},
\newblock Quantum work and heat exchanged during sudden quenches of strongly
  coupled systems (to be published).

\bibitem{rivas2019refined}
{\'A}.~Rivas,
\newblock Entropy {\bf 21}, 725 (2019).

\bibitem{campisi2009thermodynamics}
M.~Campisi, P.~Talkner, and P.~H{\"a}nggi,
\newblock Journal of Physics A: Mathematical and Theoretical {\bf 42}, 392002
  (2009).

\bibitem{buividovich2008entanglement}
{Buividovich, P.V. and Polikarpov, M.I.},
\newblock Physics Letters B {\bf 670}, 141 (2008).

\bibitem{donnelly2012decomposition}
W.~Donnelly,
\newblock Physical Review D {\bf 85}, 085004 (2012).

\bibitem{casini2014remarks}
H.~Casini, M.~Huerta, and J.~A. Rosabal,
\newblock Physical Review D {\bf 89}, 085012 (2014).

\bibitem{radicevic2014notes}
D.~Radicevic,
\newblock arXiv preprint arXiv:1404.1391  (2014).

\bibitem{aoki2015definition}
S.~Aoki {\em et~al.},
\newblock Journal of High Energy Physics {\bf 2015}, 187 (2015).

\bibitem{soni2016aspects}
R.~M. Soni and S.~P. Trivedi,
\newblock Journal of High Energy Physics {\bf 2016}, 136 (2016).

\bibitem{van2016entanglement}
K.~Van~Acoleyen {\em et~al.},
\newblock Physical Review Letters {\bf 117}, 131602 (2016).

\bibitem{bulgarelli2024duality}
A.~Bulgarelli and M.~Panero,
\newblock Journal of High Energy Physics {\bf 2024}, 41 (2024).

\bibitem{mitra2018quantum}
A.~Mitra,
\newblock Annual Review of Condensed Matter Physics {\bf 9}, 245 (2018).

\bibitem{altman2021quantum}
E.~Altman {\em et~al.},
\newblock PRX Quantum {\bf 2}, 017003 (2021).

\bibitem{daley2022practical}
A.~J. Daley {\em et~al.},
\newblock Nature {\bf 607}, 667 (2022).

\bibitem{Note1}
Relationships between work distributions and phase transitions were explored
  recently in the context of spin systems~\cite
  {lin2023work,kiely2023entropy,zhang2023excited,zawadzki2020work,wang2017probing,
  mascarenhas2014work,mzaouali2021work,fei2020work}.

\bibitem{li2008entanglement}
H.~Li and F.~D.~M. Haldane,
\newblock Phys. Rev. Lett. {\bf 101}, 010504 (2008).

\bibitem{dalmonte2022entanglement}
M.~Dalmonte, V.~Eisler, M.~Falconi, and B.~Vermersch,
\newblock Annalen der Physik {\bf 534}, 2200064 (2022).

\bibitem{kokail2021entanglement}
C.~Kokail, R.~van Bijnen, A.~Elben, B.~Vermersch, and P.~Zoller,
\newblock Nature Physics {\bf 17}, 936 (2021).

\bibitem{kokail2021quantum}
C.~Kokail {\em et~al.},
\newblock Phys. Rev. Lett. {\bf 127}, 170501 (2021).

\bibitem{joshi2023exploring}
M.~K. Joshi {\em et~al.},
\newblock Nature {\bf 624}, 539 (2023).

\bibitem{mueller2023quantum}
N.~Mueller {\em et~al.},
\newblock PRX Quantum {\bf 4}, 030323 (2023).

\bibitem{jarzynski2004nonequilibrium}
C.~Jarzynski,
\newblock Journal of Statistical Mechanics: Theory and Experiment {\bf 2004},
  P09005 (2004).

\bibitem{talkner2020colloquium}
P.~Talkner and P.~H{\"a}nggi,
\newblock Reviews of Modern Physics {\bf 92}, 041002 (2020).

\bibitem{Note2}
Boltzmann's constant is set to one throughout the paper.

\bibitem{eisert2015quantum}
J.~Eisert, M.~Friesdorf, and C.~Gogolin,
\newblock Nature Physics {\bf 11}, 124 (2015).

\bibitem{bernien2017probing}
H.~Bernien {\em et~al.},
\newblock Nature {\bf 551}, 579 (2017).

\bibitem{zhang2017observation}
J.~Zhang {\em et~al.},
\newblock Nature {\bf 551}, 601 (2017).

\bibitem{tan2021domain}
W.~L. Tan {\em et~al.},
\newblock Nature Physics {\bf 17}, 742 (2021).

\bibitem{ebadi2021quantum}
S.~Ebadi {\em et~al.},
\newblock Nature {\bf 595}, 227 (2021).

\bibitem{de2023non}
A.~De {\em et~al.},
\newblock arXiv preprint arXiv:2309.10856  (2023).

\bibitem{martinez2016real}
E.~A. Martinez {\em et~al.},
\newblock Nature {\bf 534}, 516 (2016).

\bibitem{klco2018quantum}
N.~Klco {\em et~al.},
\newblock Physical Review A {\bf 98}, 032331 (2018).

\bibitem{nguyen2022digital}
N.~H. Nguyen {\em et~al.},
\newblock PRX Quantum {\bf 3}, 020324 (2022).

\bibitem{kharzeev2020real}
D.~E. Kharzeev and Y.~Kikuchi,
\newblock Physical Review Research {\bf 2}, 023342 (2020).

\bibitem{zhou2022thermalization}
Z.-Y. Zhou {\em et~al.},
\newblock Science {\bf 377}, 311 (2022).

\bibitem{Note3}
For instance, the $\lambda _k$ statistics imply whether a system behaves
  ergodically~\cite
  {geraedts2016many,yang2017entanglement,chang2019evolution,rakovszky2019signatures,mueller2022thermalization}.
  (Ergodicity is necessary for thermalization.) Also, a gap in the
  $H_S^{\protect \rm ent}$ spectrum may signal a topologically ordered
  phase~\cite
  {li2008entanglement,mueller2022thermalization,zache2022entanglement}.

\bibitem{elben2019statistical}
A.~Elben, B.~Vermersch, C.~F. Roos, and P.~Zoller,
\newblock Physical Review A {\bf 99}, 052323 (2019).

\bibitem{huang2020predicting}
H.-Y. Huang, R.~Kueng, and J.~Preskill,
\newblock Nature Physics {\bf 16}, 1050 (2020).

\bibitem{huang2022quantum}
H.-Y. Huang {\em et~al.},
\newblock Science {\bf 376}, 1182 (2022).

\bibitem{elben2023randomized}
A.~Elben {\em et~al.},
\newblock Nature Reviews Physics {\bf 5}, 9 (2023).

\bibitem{pichler2016measurement}
H.~Pichler, G.~Zhu, A.~Seif, P.~Zoller, and M.~Hafezi,
\newblock Physical Review X {\bf 6}, 041033 (2016).

\bibitem{dalmonte2018quantum}
M.~Dalmonte, B.~Vermersch, and P.~Zoller,
\newblock Nature Physics {\bf 14}, 827 (2018).

\bibitem{bringewatt2023randomized}
J.~Bringewatt, J.~Kunjummen, and N.~Mueller,
\newblock {Quantum} {\bf 8}, 1300 (2024).

\bibitem{mueller2024notes}
N.~Mueller,
\newblock Unpublished notes, 2024.

\bibitem{matsumoto2022calculation}
K.~Matsumoto {\em et~al.},
\newblock Japanese Journal of Applied Physics {\bf 61}, 042002 (2022).

\bibitem{wu2022estimating}
Y.~Wu and J.~B. Wang,
\newblock Quantum Science and Technology {\bf 7}, 025006 (2022).

\bibitem{arunachalam2022simpler}
S.~Arunachalam, V.~Havlicek, G.~Nannicini, K.~Temme, and P.~Wocjan,
\newblock Quantum {\bf 6}, 789 (2022).

\bibitem{bravyi2021complexity}
S.~Bravyi, A.~Chowdhury, D.~Gosset, and P.~Wocjan,
\newblock Nature Physics {\bf 18}, 1367 (2022).

\bibitem{kawai2007dissipation}
R.~Kawai, J.~M.~R. Parrondo, and C.~Van~den Broeck,
\newblock Phys. Rev. Lett. {\bf 98}, 080602 (2007).

\bibitem{jarzynski2011equalities}
C.~Jarzynski,
\newblock Annual Review of Condensed Matter Physics {\bf 2}, 329  (2011).

\bibitem{Note4}
In the Standard Model of particle physics, nature sets the value of $g$.

\bibitem{stryker2019oracles}
J.~R. Stryker,
\newblock Physical Review A {\bf 99}, 042301 (2019).

\bibitem{raychowdhury2020solving}
I.~Raychowdhury and J.~R. Stryker,
\newblock Physical Review Research {\bf 2}, 033039 (2020).

\bibitem{zohar2011confinement}
E.~Zohar and B.~Reznik,
\newblock Phys. Rev. Lett. {\bf 107}, 275301 (2011).

\bibitem{banerjee2012atomic}
D.~Banerjee {\em et~al.},
\newblock Phys. Rev. Lett. {\bf 109}, 175302 (2012).

\bibitem{zohar2013simulating}
E.~Zohar, J.~I. Cirac, and B.~Reznik,
\newblock Phys. Rev. Lett. {\bf 110}, 055302 (2013).

\bibitem{tagliacozzo2013simulation}
L.~Tagliacozzo, A.~Celi, P.~Orland, M.~W. Mitchell, and M.~Lewenstein,
\newblock Nature Communications {\bf 4}, 2615 (2013).

\bibitem{hauke2013quantum}
P.~Hauke, D.~Marcos, M.~Dalmonte, and P.~Zoller,
\newblock Phys. Rev. X {\bf 3}, 041018 (2013).

\bibitem{davoudi2023towards}
Z.~Davoudi, N.~Mueller, and C.~Powers,
\newblock Physical Review Letters {\bf 131}, 081901 (2023).

\bibitem{halimeh2021gauge}
J.~C. Halimeh, H.~Lang, J.~Mildenberger, Z.~Jiang, and P.~Hauke,
\newblock PRX Quantum {\bf 2}, 040311 (2021).

\bibitem{halimeh2022stabilizing}
J.~C. Halimeh and P.~Hauke,
\newblock arXiv preprint arXiv:2204.13709  (2022).

\bibitem{SupplementalMaterial}
See Supplemental Material for discussions regarding numerical stability of
  results and the use of weak-coupling approximations in calculating
  thermodynamic quantities for the LGT system described in the main text. This
  Supplemental Material contains Ref.~\cite{desa2020low}.

\bibitem{Note5}
The mass term's sign alternates between even-index and odd-index sites,
  consistently with the staggering of fermionic matter~\cite
  {kogut1975hamiltonian}. Despite working with a bosonic theory, we keep the
  staggering convention.

\bibitem{Note6}
For a pure gauge theory, which lacks matter fields, the operator reduces to
  $G_n\mathrel {\mathop :}\mathrel {\mkern -1.2mu}=\protect \tilde {\sigma
  }^x_n \protect \tilde {\sigma }^x_{n-1}$. This is the familiar generator of a
  local $\protect \mathbb {Z}_2$ symmetry.

\bibitem{Note7}
$G_n$ has only two eigenvalues, $\pm 1$. Physically, when $g=1$, if no matter
  (or antimatter) occupies site $n$, the electric field on the left equals that
  on the right. If site $n$ contains matter (or antimatter), the electric-field
  eigenvalues on the left and right have opposite signs.

\bibitem{Note8}
Numerical computations of this work are performed by manually restricting to
  the physical Hilbert space.

\bibitem{Note9}
$\Sigma $ has the form that the chiral condensate would have if the gauge field
  were coupled to staggered fermions~\cite {kogut1975hamiltonian} (a
  Jordan-Wigner transformation maps the fermion operators to Pauli operators).
  Such a chiral condensate serves as an order parameter for chiral phase
  transitions.

\bibitem{Note10}
We have not performed a large-system analysis to confirm if the apparent phase
  transition persists in the thermodynamic limit. However, large-system
  analyses of a similar model (with the same discrete chiral symmetry as ours)
  reveal two phases, confined and deconfined~\cite
  {kebrivc2021confinement,kebrivc2023confinement}. While these studies do not
  directly probe the spontaneous breaking of the discrete chiral symmetry, the
  symmetry is broken in the confined phase.

\bibitem{wen1990topological}
X.-G. Wen,
\newblock International Journal of Modern Physics B {\bf 4}, 239 (1990).

\bibitem{kitaev2003fault}
{Kitaev, A.Yu.},
\newblock Annals of Physics {\bf 303}, 2 (2003).

\bibitem{levin2005string}
M.~A. Levin and X.-G. Wen,
\newblock Physical Review B {\bf 71}, 045110 (2005).

\bibitem{bernevig2013topological}
{Bernevig, B. A. (with T.L. Hughes)},
\newblock {\em {Topological Insulators and Topological Superconductors}}
  (Princeton University Press, Princeton, NJ, 2013).

\bibitem{seo2013thermodynamic}
K.~Seo, C.~Zhang, and S.~Tewari,
\newblock Physical Review A {\bf 87}, 063618 (2013).

\bibitem{kempkes2016universalities}
{Kempkes, S. N. and Quelle, A and Morais Smith, C.},
\newblock Scientific Reports {\bf 6}, 38530 (2016).

\bibitem{caputo2018topological}
D.~Caputo {\em et~al.},
\newblock Nature Materials {\bf 17}, 145 (2018).

\bibitem{carollo2020geometry}
A.~Carollo, D.~Valenti, and B.~Spagnolo,
\newblock Physics Reports {\bf 838}, 1 (2020).

\bibitem{heyl2018dynamical}
M.~Heyl,
\newblock Reports on Progress in Physics {\bf 81}, 054001 (2018).

\bibitem{schweizer2019floquet}
C.~Schweizer {\em et~al.},
\newblock Nature Physics {\bf 15}, 1168 (2019).

\bibitem{yang2020observation}
B.~Yang {\em et~al.},
\newblock Nature {\bf 587}, 392 (2020).

\bibitem{notarnicola2020real}
S.~Notarnicola, M.~Collura, and S.~Montangero,
\newblock Physical Review Research {\bf 2}, 013288 (2020).

\bibitem{surace2020lattice}
F.~M. Surace {\em et~al.},
\newblock Physical Review X {\bf 10}, 021041 (2020).

\bibitem{klco20202}
N.~Klco, M.~J. Savage, and J.~R. Stryker,
\newblock Physical Review D {\bf 101}, 074512 (2020).

\bibitem{ciavarella2021trailhead}
A.~Ciavarella, N.~Klco, and M.~J. Savage,
\newblock Phys. Rev. D {\bf 103}, 094501 (2021).

\bibitem{de2022quantum}
W.~A. de~Jong {\em et~al.},
\newblock Physical Review D {\bf 106}, 054508 (2022).

\bibitem{mildenberger2022probing}
J.~Mildenberger, W.~Mruczkiewicz, J.~C. Halimeh, Z.~Jiang, and P.~Hauke,
\newblock arXiv preprint arXiv:2203.08905  (2022).

\bibitem{halimeh2022tuning}
J.~C. Halimeh, I.~P. McCulloch, B.~Yang, and P.~Hauke,
\newblock PRX Quantum {\bf 3}, 040316 (2022).

\bibitem{zhang2023observation}
W.-Y. Zhang {\em et~al.},
\newblock arXiv preprint arXiv:2306.11794  (2023).

\bibitem{charles2024simulating}
C.~Charles {\em et~al.},
\newblock Physical Review E {\bf 109}, 015307 (2024).

\bibitem{surace2021scattering}
F.~M. Surace and A.~Lerose,
\newblock New Journal of Physics {\bf 23}, 062001 (2021).

\bibitem{belyansky2023high}
R.~Belyansky {\em et~al.},
\newblock Phys. Rev. Lett. {\bf 132}, 091903 (2024).

\bibitem{chai2023entanglement}
Y.~Chai {\em et~al.},
\newblock arXiv preprint arXiv:2312.02272  (2023).

\bibitem{su2024cold}
G.-X. Su, J.~Osborne, and J.~C. Halimeh,
\newblock arXiv preprint arXiv:2401.05489  (2024).

\bibitem{farrell2024quantum}
R.~C. Farrell, M.~Illa, A.~N. Ciavarella, and M.~J. Savage,
\newblock Phys. Rev. D {\bf 109}, 114510 (2024).

\bibitem{davoudi2024scattering}
Z.~Davoudi, C.-C. Hsieh, and S.~V. Kadam,
\newblock {Quantum} {\bf 8}, 1520 (2024).

\bibitem{lin2023work}
F.-L. Lin and C.-Y. Huang,
\newblock Phys. Rev. Res. {\bf 6}, 023169 (2024).

\bibitem{kiely2023entropy}
A.~Kiely, E.~O'Connor, T.~Fogarty, G.~T. Landi, and S.~Campbell,
\newblock Phys. Rev. Res. {\bf 5}, L022010 (2023).

\bibitem{zhang2023excited}
H.~Zhang, Y.~Qian, Z.-X. Niu, and Q.~Wang,
\newblock Phys. Rev. E {\bf 109}, 064110 (2024).

\bibitem{zawadzki2020work}
K.~Zawadzki, R.~M. Serra, and I.~D'Amico,
\newblock Physical Review Research {\bf 2}, 033167 (2020).

\bibitem{wang2017probing}
Q.~Wang and H.~T. Quan,
\newblock Physical Review E {\bf 96}, 032142 (2017).

\bibitem{mascarenhas2014work}
E.~Mascarenhas {\em et~al.},
\newblock Physical Review E {\bf 89}, 062103 (2014).

\bibitem{mzaouali2021work}
Z.~Mzaouali, R.~Puebla, J.~Goold, M.~El~Baz, and S.~Campbell,
\newblock Physical Review E {\bf 103}, 032145 (2021).

\bibitem{fei2020work}
Z.~Fei, N.~Freitas, V.~Cavina, H.~T. Quan, and M.~Esposito,
\newblock Phys. Rev. Lett. {\bf 124}, 170603 (2020).

\bibitem{geraedts2016many}
S.~D. Geraedts, R.~Nandkishore, and N.~Regnault,
\newblock Physical Review B {\bf 93}, 174202 (2016).

\bibitem{yang2017entanglement}
Z.-C. Yang, A.~Hamma, S.~M. Giampaolo, E.~R. Mucciolo, and C.~Chamon,
\newblock Physical Review B {\bf 96}, 020408 (2017).

\bibitem{chang2019evolution}
P.-Y. Chang, X.~Chen, S.~Gopalakrishnan, and J.~Pixley,
\newblock Physical Review Letters {\bf 123}, 190602 (2019).

\bibitem{rakovszky2019signatures}
T.~Rakovszky, S.~Gopalakrishnan, S.~Parameswaran, and F.~Pollmann,
\newblock Physical Review B {\bf 100}, 125115 (2019).

\bibitem{mueller2022thermalization}
N.~Mueller, T.~V. Zache, and R.~Ott,
\newblock Physical Review Letters {\bf 129}, 011601 (2022).

\bibitem{zache2022entanglement}
T.~V. Zache, C.~Kokail, B.~Sundar, and P.~Zoller,
\newblock Quantum {\bf 6}, 702 (2022).

\bibitem{desa2020low}
C.~De~Sa,
\newblock CS4787 Lecture 21, Cornell University  (2020).

\bibitem{kebrivc2021confinement}
M.~Kebri{\v{c}}, L.~Barbiero, C.~Reinmoser, U.~Schollw{\"o}ck, and F.~Grusdt,
\newblock Physical Review Letters {\bf 127}, 167203 (2021).

\bibitem{kebrivc2023confinement}
M.~Kebri{\v{c}}, J.~C. Halimeh, U.~Schollw\"ock, and F.~Grusdt,
\newblock Phys. Rev. B {\bf 109}, 245110 (2024).

\end{thebibliography}

\newpage 
\onecolumngrid

\section*{Supplemental Material:
\\
``Quantum thermodynamics of nonequilibrium processes in lattice gauge theories''}

\subsection{Supplemental Note I:
\\
Convergence of thermodynamic quantities with $\beta$ 
\\
in the lattice-gauge-theory example}\label{appendix:betaconvergence}

\noindent
The main text reports on simulations that require care because numerical instabilities can arise from multiple sources.
Matrix exponentials $e^{-\beta H}$ can have eigenvalues with very large or very small magnitudes, especially at large $\beta$ values. Such magnitudes can lead to instabilities in the numerical evaluation of $e^{-\beta H}$. Instabilities can arise also in matrix-logarithm calculations, due to e.g., large or small eigenvalues. Large eigenvalues destabilize calculations more than small eigenvalues do, because the absolute error in a floating-point representation increases with the magnitude of the number represented ~\cite{desa2020low}. 

Therefore, we prioritize the elimination of large eigenvalues from matrix exponentials. We add constant shifts to the Hamiltonians $H_\sur(t=0^-)$ and $H_\sur(t=0^+)$; see Eq. (13). The shifts manifest as uniform on-site energy densities of magnitude $c$. We choose $c$ such that $H_S(t)$, $H_R$, and $H_\sur(t)$ have only non-negative eigenvalues at the system sizes studied; hence exponentials $e^{-\beta H}$ lack large eigenvalues. The energy-density terms are partitioned between the sub-Hamiltonians according to Eq. (1) like the other on-site terms (shown in Fig. 2). Such constant shifts cannot change physical observables' qualitative behaviors.

\begin{figure*}[h]
    \centering
    \includegraphics[width=0.75\linewidth]{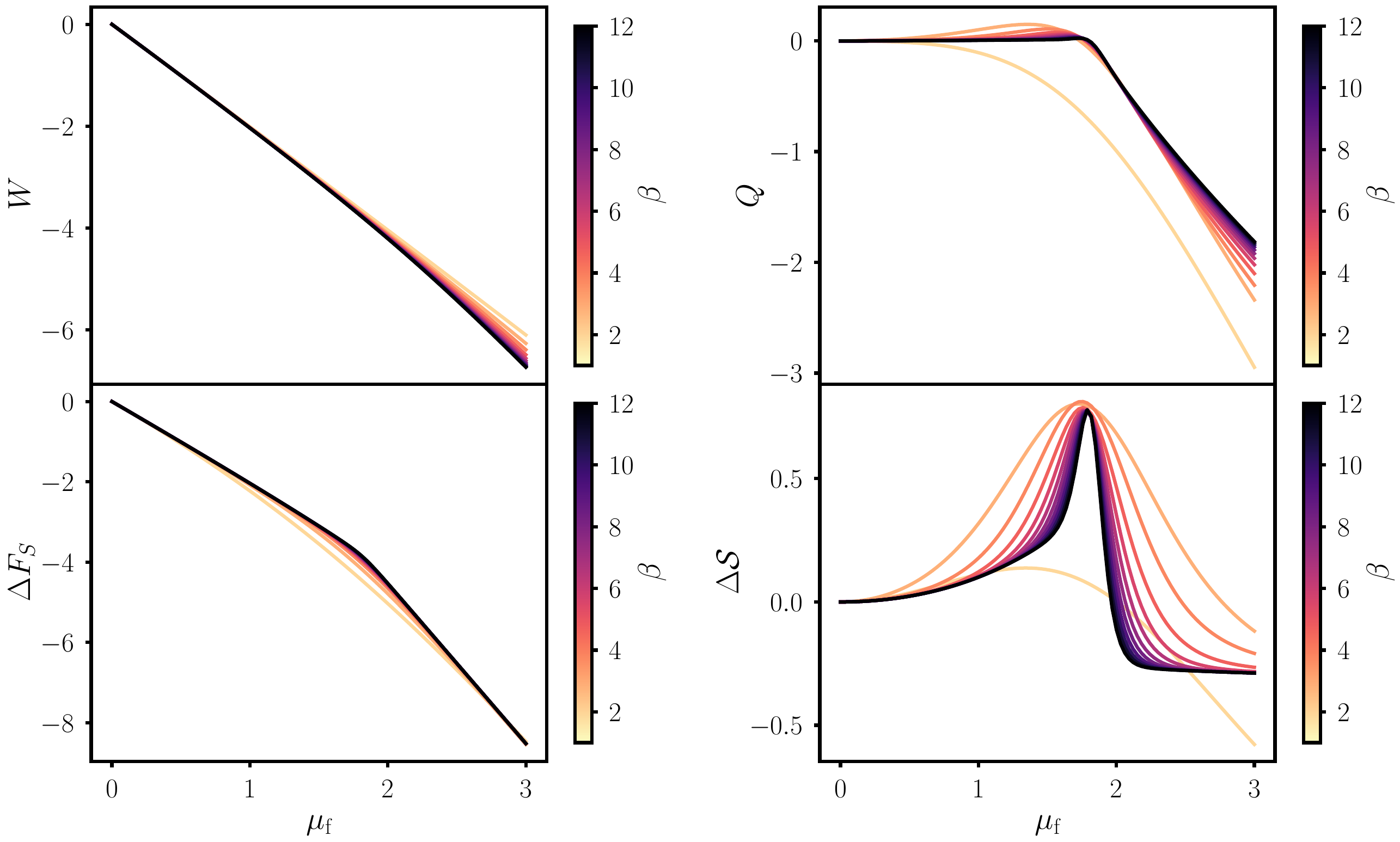}
    \caption{Convergence of thermodynamic quantities with increasing $\beta$ for the quench of $\mu_{\rm i}=0$ to $\mu_{\rm f} > 0$, as in the main text.
    }
    \label{fig:1D_muquench_HMF-based_thermo_quantities_beta_sweep_fig}
\end{figure*}

\begin{figure*}[h]
    \centering
    \includegraphics[width=0.4\linewidth]{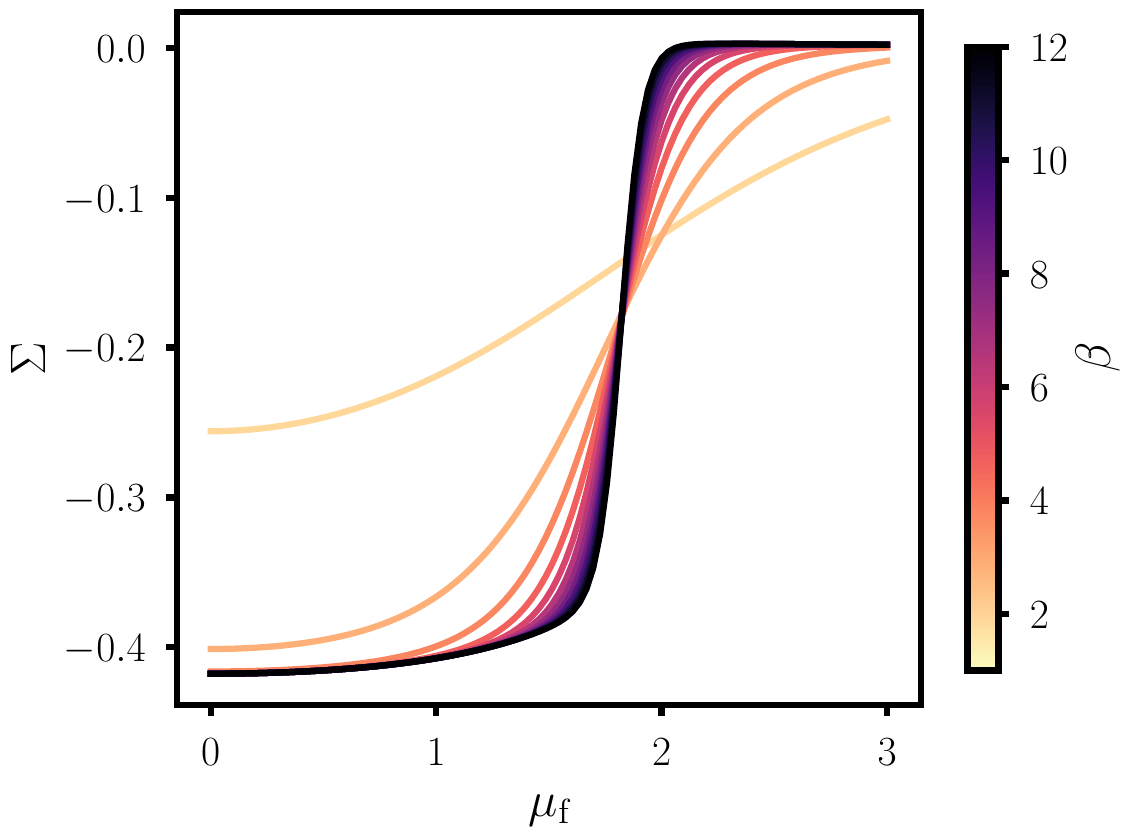}
    \caption{Convergence of $\Sigma$ with increasing $\beta$ for the quench of $\mu_{\rm i}=0$ to $\mu_{\rm f} > 0$, as in the main text.
    }
    \label{fig:1D_muquench_Sigma_beta_sweep_fig}
\end{figure*}

Numerical errors may still occur when Hamiltonians have large positive eigenvalues. Thus, we test the numerical stability of the results shown in Fig. 3. We investigate whether thermodynamic quantities, as well as the chiral condensate, behave smoothly as functions of $\beta$. Figures~\ref{fig:1D_muquench_HMF-based_thermo_quantities_beta_sweep_fig} and \ref{fig:1D_muquench_Sigma_beta_sweep_fig} display these quantities. They exhibit smooth behaviors as $\beta$ varies, indicating the absence of instabilities. Evidence of the phase transition becomes sharper at smaller temperatures (larger $\beta$ values). The main-text plots correspond to $\beta=10$.
%

\subsection{Supplemental Note II:
\\
Weak-coupling quantum thermodynamics applied to 
\\
the lattice-gauge-theory example}
\noindent
Strong-coupling quantum thermodynamics becomes relevant when the system-reservoir interaction contributes non-negligibly to the system's internal energy. The penalty terms in LGTs, introduced in Eq. (12) to impose Gauss's laws, act at all sites. Therefore, the boundary penalty terms' contribution may not appear significant, compared to the 
contribution of the terms acting on the system's bulk, particularly if the number of system DOFs exceeds the number of DOFs at the boundary. A deeper perspective, however, is that local constraints imposed at all sites lead to highly nonlocal structure in the states. Hence, we might expect the system's internal energy to depend on reservoir degrees of freedom.

We now confirm that the example in the main text is in the strong-coupling regime. The example is a $Z_2$ LGT coupled to spin-$\tfrac{1}{2}$ bosonic matter in (1+1)D with $N=6$ sites in $S \cup R$ and $N_S=4$ sites in $S$. We evaluate $|\langle V_\sur \rangle / \langle H_S(t) \rangle|$ in our quench protocol's initial and final (equilibrium) states. In both cases, the ratio is non-negligible ($\gtrsim 0.1$) at all the $\mu_f$ values used in our numerical study. We neglected contributions from the uniform on-site energy-density terms when computing $\langle H_S(t) \rangle$.

\begin{figure}[H]
    \centering
    \includegraphics[width=0.5\linewidth]{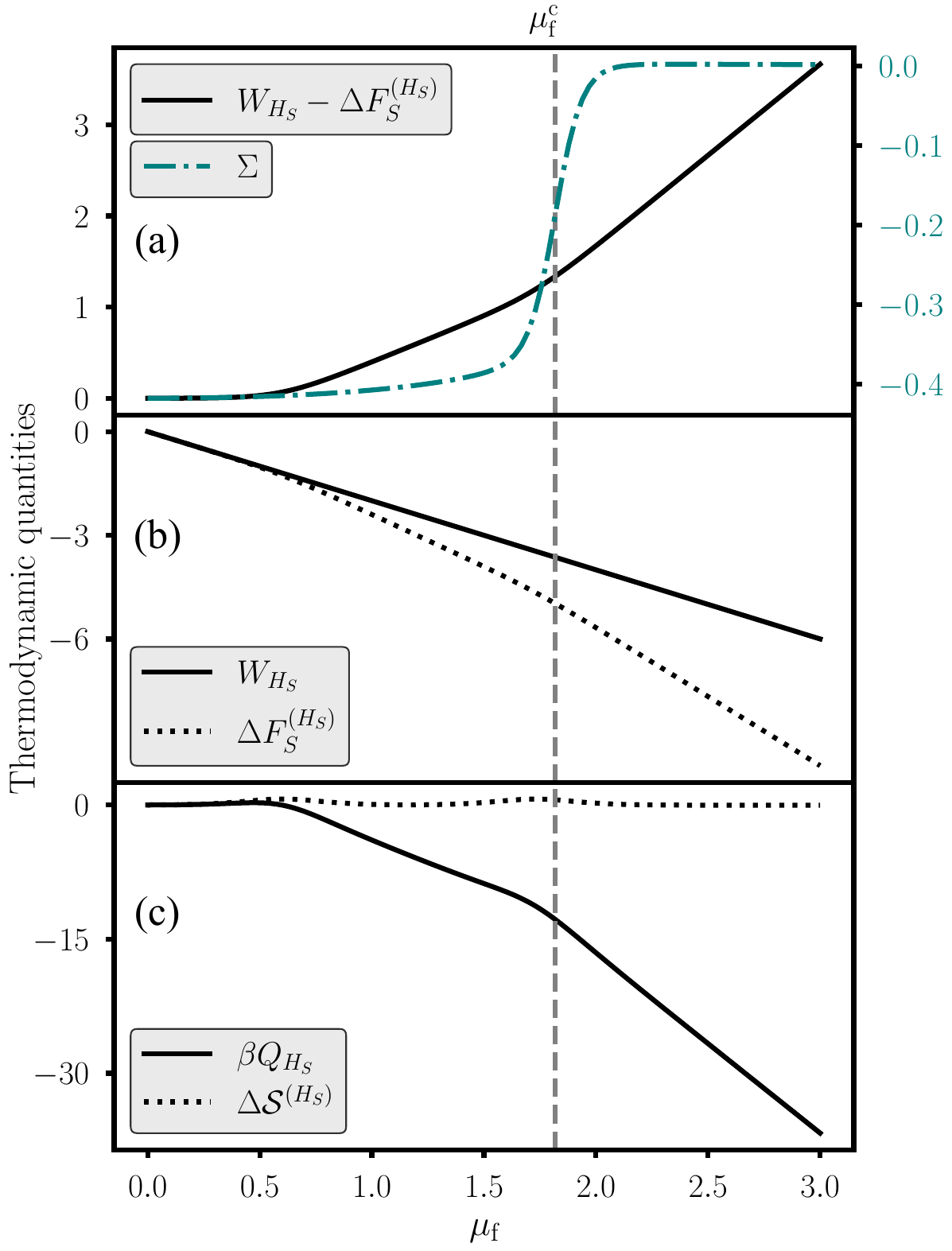}
    \caption{Computing thermodynamic quantities using weak-coupling assumptions: For illustrative purposes, we make the approximations (argued to be incorrect here) $e^{-\beta H_S^*(t)}/\Tr(e^{-\beta H_S^*(t)}) \approx e^{-\beta H_S(t)}/\Tr(e^{-\beta H_S(t)})$ and $U_S(t) \approx \langle H_S(t) \rangle$. Thermodynamic quantities are plotted as functions of the final chemical potential, $\mu_{\text{f}}$, 
    during the instantaneous quench from $\mu_{\text{i}}=0$: (a) Dissipated work $W_{\rm{diss}}$ (solid, black) and chiral condensate $\Sigma$ (dot-dashed, teal). (b) Work $W$ (solid, black) and change $\Delta F_S$ in the system's free energy (dotted, black). (c) Normalized heat $\beta Q$ (solid, black) and change $\Delta \mathcal{S}$ in the system's entropy (dotted, black). In all subfigures, we plot $\mu_\text{f}^{\rm c}$, the $\mu_{\text{f}}$ value where $\Sigma'$ is maximized (dashed, gray). For these plots, $N=6$, $N_S=4$, $J=-1/2$, $\epsilon=1/2$, $m=1/2$, $\mu_\text{i}=0$, and $\beta=10$, as in Fig. 3.
    }
    \label{fig:H_S-based quantities plot N_S=4}
\end{figure}

Which values do thermodynamic quantities assume in this LGT if one nevertheless applies weak-coupling thermodynamics? We plot in Fig.~\ref{fig:H_S-based quantities plot N_S=4} the same quantities as in Fig. 3 of the main text, but computed with weak-coupling quantum-thermodynamics definitions: the system's internal energy is equated with the expectation value of its Hamiltonian, $H_S$, as opposed to the Hamiltonian of mean force, $H_S^*$. Furthermore, the system states $\rho_S = e^{-\beta H_S^*}/\Tr(e^{-\beta H_S^*})$ become $\rho_S \approx e^{-\beta H_S}/\Tr(e^{-\beta H_S})$. This approximation affects computations of not only expectation values, but also equilibrium properties such as free energy and entropy. 

Figure~\ref{fig:H_S-based quantities plot N_S=4} shows that thermodynamic quantities' values differ from their strong-coupling counterparts. The weak-coupling quantities' behaviors near the expected phase transition are rather smooth, exhibiting diminished abilities to signal the transition. 
The chiral condensate (dashed green) is calculated from the system's true reduced state, as in the main text. 

In summary, our observations motivate the use of the strong-coupling framework to describe LGTs' open-quantum-system thermodynamics.

\end{document}